\documentstyle[twoside,fleqn,espcrc2,psfig]{article}

\newcommand{\AmS}{{\protect\the\textfont2
  A\kern-.1667em\lower.5ex\hbox{M}\kern-.125emS}}

\title{2d quantum gravity with discrete edge lengths
      \thanks{Supported in part by Fonds zur F\"orderung der Wissenschaftlichen
              Forschung under Contract P11141-PHY.}}

\author{E. Bittner, H. Markum and J. Riedler
\address{Institut f\"ur Kernphysik,
        Technische Universit\"at Wien,
        Wiedner Hauptstr. 8-10, 1040 Vienna, Austria}}

\begin{document}

\begin{abstract}
An approximation of the Standard Regge Calculus (SRC) was proposed by the 
$Z_2$-Regge Model ($Z_2$RM). There the edge lengths of the simplicial complexes
are restricted to only two possible values, both always compatible with the
triangle inequalities.  To examine the effect of discrete edge lengths, we
define two models to describe the transition from the $Z_2$RM to the SRC.
These models allow to choose the number of possible link lengths to be
$n = \{ 4,8,16,32,64,\dots \}$ and differ mainly in the scaling of the
quadratic link lengths. The first extension, the $X^1_n$-Model, keeps the
edge lengths limited and still behaves rather similar to the "spin-like"
$Z_2$RM. The vanishing critical cosmological constant is reproduced by the
second extension, the $X^C_n$-Model, which allows for increasing edge lengths.
In addition the area expectation values are consistent with the scaling
relation of the SRC.
\end{abstract}

\maketitle

The approach described in this work relies on Regge's approach to gravitation
\cite{regge}, a discrete description of general relativity. The theory is
regularized through the introduction of a natural cutoff, the lattice spacing,
to approximate space-time by a simplicial lattice. The lattice thus becomes a
dynamical object, with the squared edge lengths $q$ describing the evolution of
space-time. This so-called Standard Regge Calculus (SRC) provides an interesting
method to explore quantum gravity in a non-perturbative way, see \cite{loll}
for a comprehensive review.
Quantization of 2d SRC proceeds by evaluating the path integral
\begin{equation}\label{Z}
Z = \prod_l \int \frac{dq_l}{q_l^m} {\cal F}(q_l) 
    \exp(k\mbox{$\sum\limits_s$}\delta_s -\lambda\mbox{$\sum\limits_t$}A_t)~.
\end{equation}
In principle the functional integration should extend over all geometries on
all possible topologies, but, as is usually done, we restrict ourselves to 
one specific topology, the torus.
${\cal F}$ is a function to assure that only Euclidean configurations of 
links may contribute and the real parameter $m$ allows to choose a particular
measure. The action in the exponential of (\ref{Z}) consists of the sum
over the site-associated deficit angles $\delta_s$ times the
gravitational coupling $k$, that is the Regge-Einstein term, and 
the sum over all triangle areas $A_t$ times the cosmological constant $\lambda$.
In two dimensions the Regge-Einstein action gives the Euler characteristic
of the surface and may be dropped for fixed topology.

Although the SRC code can be efficiently vectorized for large scale computing, 
the simulations are still a very time demanding enterprise. One therefore 
seeks for suitable
approximations which will simplify the SRC and yet retain most of its universal
features. The $Z_2$-Regge Model ($Z_2$RM) \cite{juri} could be such a desired
simplification.
Here the quadratic link lengths of the simplicial complexes are allowed to take
on only the two values
\begin{equation}\label{ql}
q_l = 1 + \epsilon \sigma_l~,\quad \sigma_l = \pm 1~.
\end{equation}
Then the area of a triangle $t$ with edges
$q_1,q_2,q_l$ can be expressed as
\begin{eqnarray} \label{at}
A_t&\!\!\!=&\!\!\!c_0+c_1(\sigma_1+\sigma_2+\sigma_l)+
c_2(\sigma_1\sigma_2 +\nonumber\\
&&\!\!\!+\;\sigma_1\sigma_l+\sigma_2\sigma_l)+c_3\sigma_1\sigma_2\sigma_l ~.
\end{eqnarray}
The coefficients $c_i$ depend on $\epsilon$ only and impose the condition
$\epsilon <\frac{3}{5}=\epsilon_{max}$ in order to have real and positive
triangle areas, i.e. \hbox{${\cal F}=1$} for all possible link length
configurations. This is quite different to SRC where many potential updates
either violate the triangle inequality or the manifold property. The action
as well as the measure in the path integral (\ref{Z}) can be rewritten
in terms of the $Z_2$-variables $\sigma_l$, see \cite{juri} for more details.
Moreover, if we choose the measure parameter as
\begin{equation}
m=-\frac{c_1\lambda}{2\sum_{i=1}^\infty \frac{\epsilon^{2i-1}}{2i-1}}~,
\end{equation}
the partition function of the $Z_2$RM takes on a particularly simple form
\begin{displaymath}
Z=\!\sum_{\sigma_l=\pm 1}\!e^{-\lambda\mbox{\small{$\sum\limits_t
[c_2(\sigma_1\sigma_2+\sigma_1\sigma_l+\sigma_2\sigma_l)+
c_3\sigma_1\sigma_2\sigma_l]$}}}\!\!. 
\end{displaymath}
To investigate the transition from the $Z_2$RM to SRC we
allow $n$ values for the squared edge lengths $q_l$
\begin{eqnarray} \label{xql}
q_l&\!\!\! =&\!\!\! C (1+\epsilon \sigma_l)~, \quad 
 0 \leq \epsilon < \epsilon_{max}~,\\
\sigma_l&\!\!\! \in&\!\!\! \{ -(n-1),-(n-1)+2,\dots,(n-1)\}~.\nonumber
\end{eqnarray}
We now set the parameter $\epsilon = \frac{1}{n}$ and construct
$\epsilon_{max}$ by considering the ``worst case'' from the
point of view of the triangle inequalities. From that, it is straightforward
to show 
\begin{equation}
\epsilon_{max}=(\frac{3}{3n-1})~, \quad n\ge2~.
\end{equation}
What remains is to assign a role to the scaling factor $C$. The simplest
possibility is $C=1$, leading to an extended $Z_2$RM, which for
example in the case $n=8$ we call $X_8^1$-Regge Model. A disadvantage
of the $X_n^1$-Regge Model ($X_n^1$RM) is the restriction of the link lengths
to $q_l<2$. Thus, the triangle areas are prevented to increase arbitrarily, 
either. In order to regain the feature of no upper
boundary on the triangle areas as is manifest in SRC we set $C=\frac{n}{2}$,
and call the corresponding model $X_n^C$RM.

In our Monte Carlo simulations we employed 16x16 lattices with periodic
boundary conditions. Starting from an initial configuration consisting
of equilateral triangles, 30k thermalization steps were performed. This was
followed by 50k measurement steps and we actually used every $10^{th}$
for analysis. Error bars were determined by the standard jackknife method
using bins of 20 data.

We simulated the $X_n^1$RM for $n=\{2,4,8,16,$ $32,64,256\}$ and analyzed
expectation values of the area normalized to the total number of vertices.
The SRC is known to possess an ill-defined phase for negative cosmological
constants and
\begin{equation} \label{lAr}
\langle A \rangle = N_1 \frac{1-m}{\lambda}
\end{equation}
decreases inversely proportional for positive $\lambda$ ($N_1$ is the total
number of links). From the left plots in Fig.~1 it is visible that for 
increasing number $n$ of possible link lengths the critical couplings 
$\lambda_c$ approach $-4.2$. The naive expectation to reproduce the 
transition at $\lambda=0$ of SRC from a well to an ill-defined phase is not
realized for the $X_n^1$RM. For $\lambda<\lambda_c$ all the triangles assume
their minimum area as in the $Z_2$RM and do not grow unlimited like in
the SRC. $\langle A \rangle$ still exhibits a decrease for positive
cosmological constant $\lambda$ and the functions converge for increasing $n$.

We simulated the $X_n^C$RM for $n=\{4,8,16,32,$ $64,256,512\}$ and present
the results in the right plots of Fig.~1. For $\lambda<0$ the surface tries
to blow up as in the SRC, all the triangle areas obtain their maximum value.
The expectation values of the area approach the exact SRC result (\ref{lAr})
for increasing number $n$ of possible link lengths and $0<\lambda<1$. The
deviation of $\langle A\rangle$ from the SRC result in the region $\lambda>1$
is remarkable. The system moves into metastable states just as the SRC, but
there the breathing algorithm takes care of that problem. A breathing update
simply consists
of rescaling all quadratic link lengths $q$ by a factor $\zeta$, which is the
same as scaling the total area by $\zeta$. Unfortunately a similar algorithm
is not possible for the $X_n^C$RM. Contrary to the $X_n^1$RM, within the
$X_n^C$RM the critical coupling $\lambda_c$ approaches zero with increasing
$n$. From this point of view the $X_n^C$RM is certainly well suited to
approximate the SRC.

In order to check whether they have common universal features, too, we
consider the Liouville field susceptibility
\begin{equation}
\chi_\phi = \langle A \rangle [\langle \phi^2 \rangle -
                               \langle \phi \rangle^2] ~,
\end {equation}
with the Liouville field $\phi=\frac{1}{A}\sum_i \ln{A_i}$.
$A$ is the total area and $A_i$ the area element of the site $i$ \cite{ham}. 
From continuum field theory it is known that for fixed $A$ the
susceptibility scales according to
\begin{equation}
\ln\chi_{\phi}(L) \stackrel{L\to\infty}{\sim} c +
(2 - \eta_{\phi}) \ln L ~,
\end{equation}
with $L=\sqrt{A}$ and the Liouville field critical exponent $\eta_{\phi}=0$.
This has indeed been observed for SRC with the $dq/q$ scale invariant measure
and fixed area constraint \cite{ham}. It is, however, a priori not clear
whether this result will persist in 
\begin{figure*}[hth]
\label{fig1}
\centerline{\hbox{
\psfig{figure=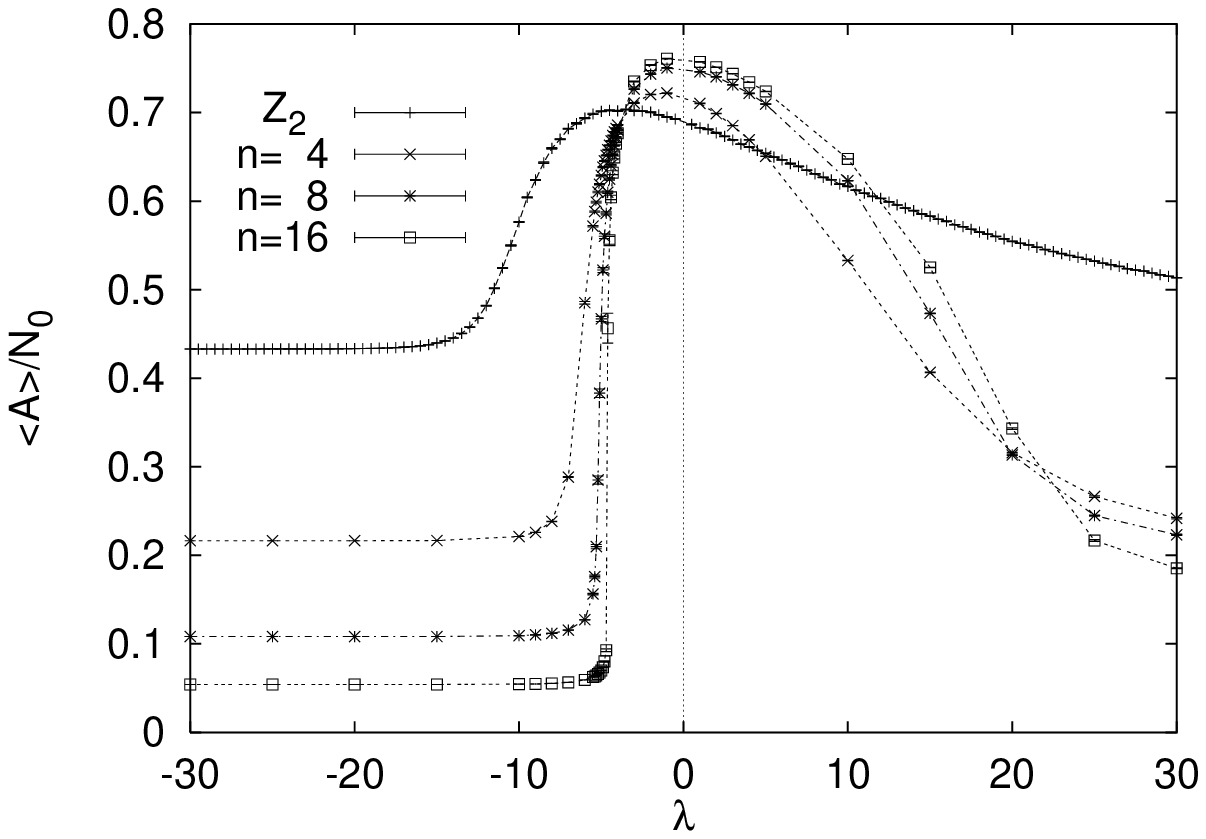,height=5.8cm,width=8cm}
\psfig{figure=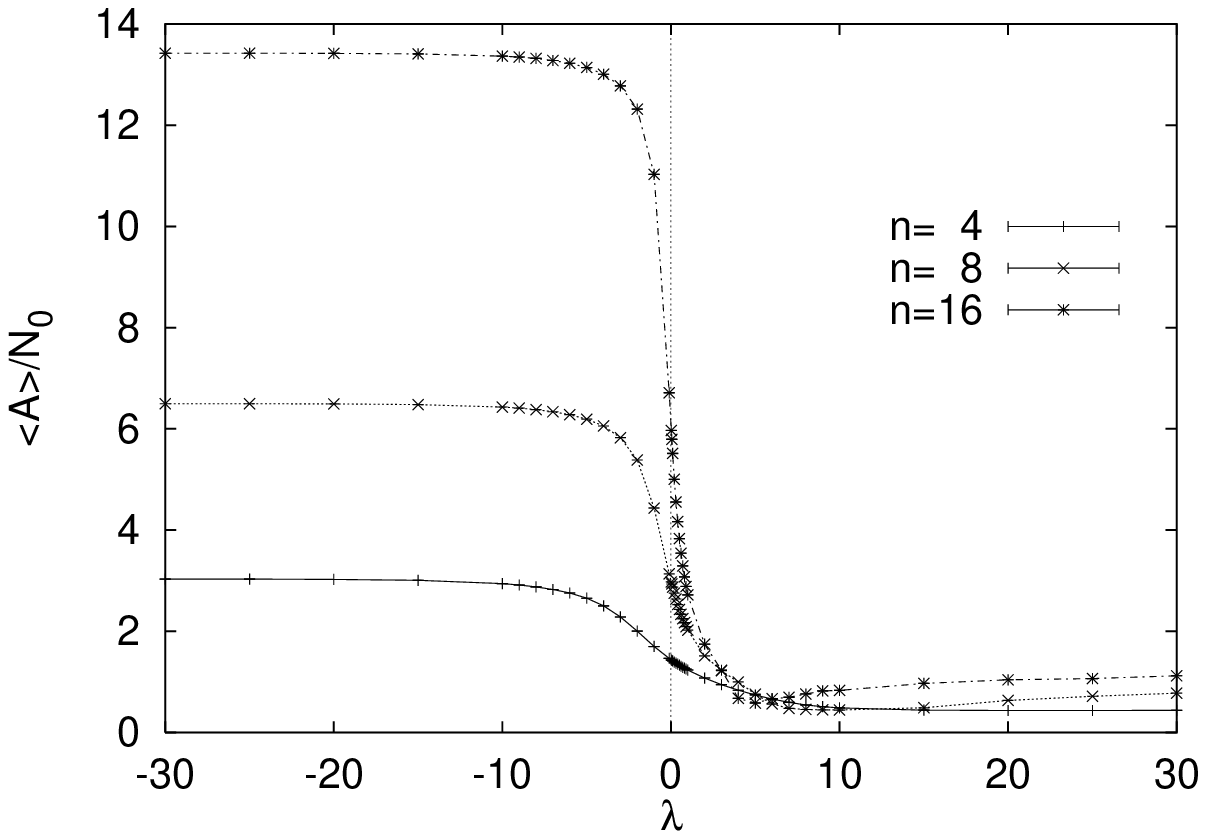,height=5.8cm,width=8cm}
}}
\centerline{\hbox{
\psfig{figure=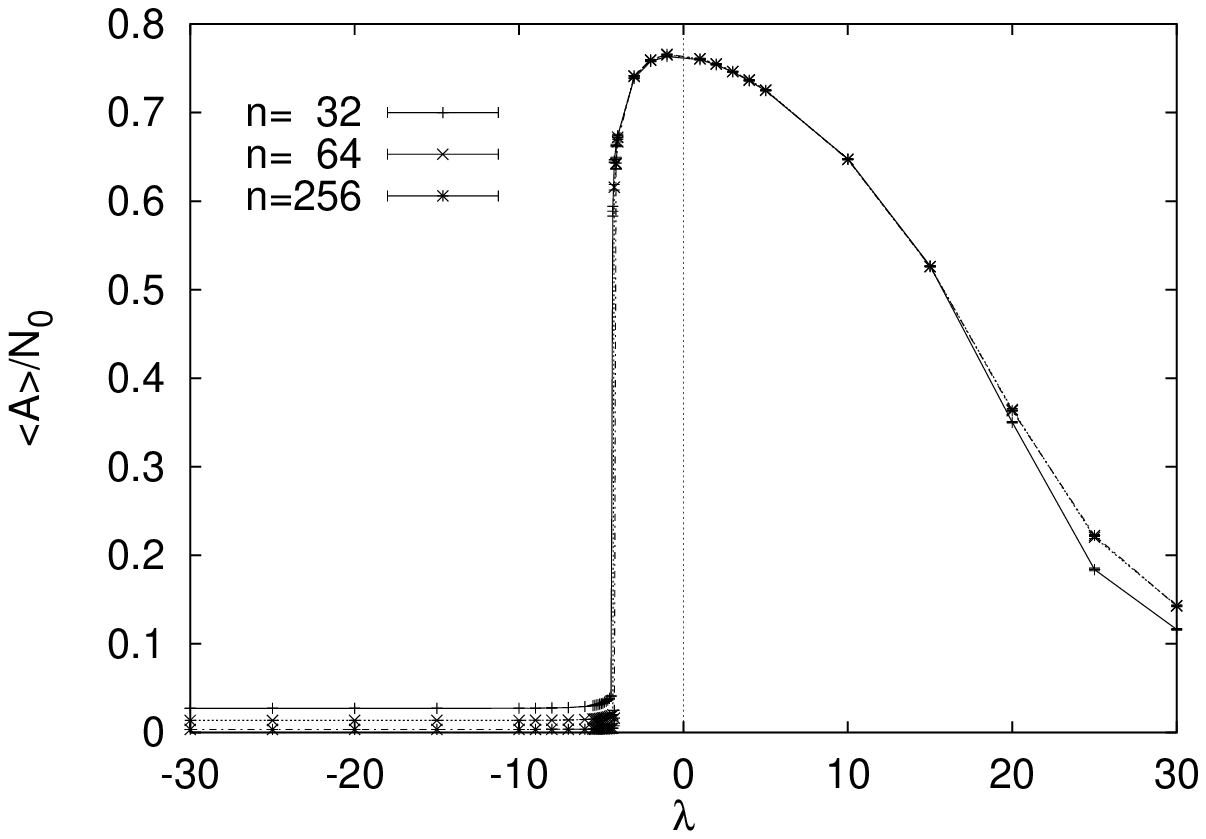,height=5.8cm,width=8cm}
\psfig{figure=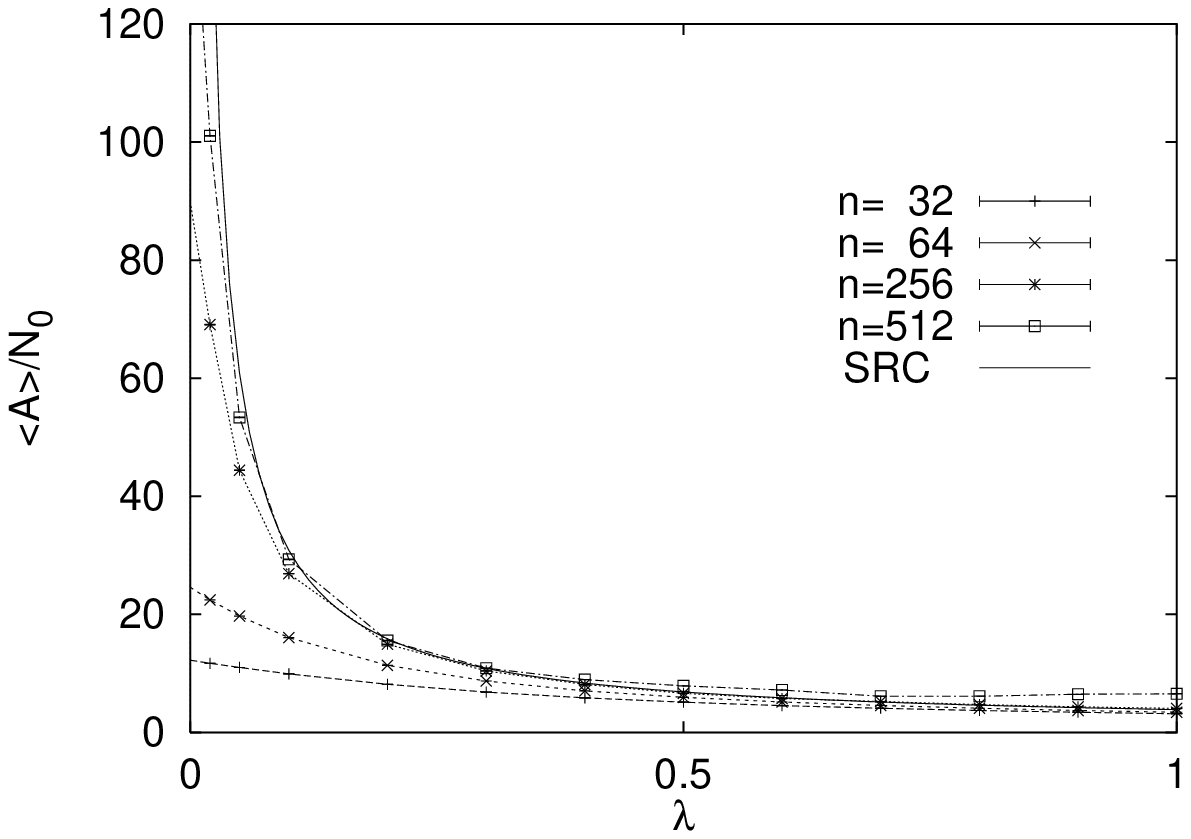,height=5.8cm,width=8cm}
}}
\vspace{-8mm}
\caption{Expectation values of the area $A$
normalized to the total number of vertices $N_0$
as a function of the cosmological constant $\lambda$ for
$X_n^1$RM (left plots) and the $X_n^C$RM (right
plots).}
\end{figure*}
the present model due to the fluctuating area and the non-scale invariant
measure. Actually we find that the Liouville field susceptibility for the
SRC and the $X_n^C$RM scales with $\eta_\phi \approx 2$.

In summary, to find discrete approximations of the SRC we constructed two
models. The first ($X_n^1$RM) has an upper limit on the link lengths,
independent of the number $n$ they are allowed to assume, whereas in the
second ($X_n^C$RM) the maximum link length increases with $n$. Thus the 
$X_n^C$RM can reproduce the phase structure of SRC and shows even the same
Liouville field critical exponent.

\end{document}